\documentclass[aps,prl,twocolumn,groupedaddress]{revtex4}

\usepackage{amsmath}

\usepackage{amssymb}

\usepackage{amsbsy}

\usepackage{graphicx}

\usepackage{color}

\begin{document}

\title{Nonlinear symmetry breaking of Aharonov-Bohm cages}

\author{Goran Gligori\'c$^{1}$, Petra P. Beli\v cev$^{1}$, Daniel Leykam$^{2}$, Aleksandra Maluckov$^{1}$}
\affiliation{$^{1}$P$^{*}$ Group, Vin\v ca Institute of Nuclear
Sciences, University of Belgrade, P.O. Box 522, 11001 Belgrade,
Serbia\\
$^{2}$Center for Theoretical Physics of Complex Systems, Institute
for Basic Science, Daejeon 34126, Republic of Korea}

\begin{abstract}
We study the influence of mean field cubic nonlinearity on
Aharonov-Bohm caging in a diamond lattice with synthetic magnetic
flux. For sufficiently weak nonlinearities the Aharonov-Bohm
caging persists as periodic nonlinear breathing dynamics. Above a
critical nonlinearity, symmetry breaking induces a sharp
transition in the dynamics and enables stronger wavepacket
spreading. This transition is distinct from other flatband
networks, where continuous spreading is induced by effective
nonlinear hopping or resonances with delocalized modes, and is in
contrast to the quantum limit, where two-particle hopping enables
arbitrarily large spreading. This nonlinear symmetry breaking
transition is readily observable in femtosecond laser-written
waveguide arrays.
\end{abstract}

\date{\today}
\maketitle

{\it Introduction.} Perfect wave localization emerges in certain
non-interacting tight binding networks via application of a
fine-tuned magnetic flux~\cite{nori,abc2d}. The localization
mechanism in such ``Aharonov-Bohm (AB) cages'' is the
flux-controlled destructive interference between different
propagation paths~\cite{abohm,abbb}, which are recombined and
forced to interfere at bottlenecks in the network. What is perhaps
most interesting about AB caging is that this perfect localization
is not limited to excitations at a precise energy (i.e. of a flat
Bloch band); it persists for arbitrary initial states. This
requires not just fine-tuning, but also a network topology
supporting closed flux-encircling plaquettes, leading to novel
topological invariants and edge
states~\cite{bergman,expzamajt,rhim2018}.

First observed in a two-dimensional ``dice'' superconducting
network nearly 20 years ago~\cite{9}, recent advances in synthetic
gauge field engineering have renewed interest in AB caging in the
context of quasi-1D and 2D mesoscopic networks including quantum
rings~\cite{fomin,p21,quantrings}, Josephson junction
arrays~\cite{doucot,doucot2005,rizzi2006,gladchenko2009,rhombus},
optical lattices~\cite{moller,junemann}, and coupled optical
waveguides~\cite{fang,Longhi,Experiment,expzamajt}, motivated by
the goal of enhancing interaction effects. In particular, single
particle (non-interacting) eigenstates in AB cages are compactly
localized but non-orthogonal, such that interactions induce
two-particle hopping processes. This destroys the caging, leading
to delocalized bound pairs, novel strongly correlated quantum
phases such as $4e$ superconductivity, and time-reversal symmetry breaking
ground states~\cite{vidal1,Vidal,takayoshi,tovmasyan,tovmasyan_arxiv,cartwright,mondaini,moller2012,correlated_review,review}.

Here we study AB caging in the presence of mean field interactions
described by the discrete nonlinear Schr\"odinger equation,
relevant to Bose-Einstein condensates~\cite{polariton_FB} and high
power light propagation in optical waveguide
arrays~\cite{heinrich}, where AB caging was very recently
observed~\cite{Experiment,expzamajt}. Due to non-orthogonality of
the compact localized eigenstates (CLS), weak nonlinearities are
already sufficient to induce linear instabilities via coupling
between neighbouring CLS~\cite{Carlo}. We demonstrate through
numerical simulations that these instabilities are weak in the
sense that the dynamics remain (quasi-)periodic and spreading to
more distant lattice sites remains negligible; most of the power
remains confined to the initially-excited CLS as a stable
breathing mode. At a critical nonlinearity strength we observe a
sharp transition in the dynamics at which this breather becomes
unstable due to nonlinear symmetry breaking between the two legs
of each plaquette. Leg-dependent nonlinear phase shifts can then
break the AB cage, leading to delocalization beyond that allowed
within the linear stability analysis (LSA). Interestingly, this
transition is not specific to the AB cage limit, but is also
robust to detunings of the effective flux, suggesting it is rooted
in the presence of bottlenecks. This nonlinearity-induced
transition may be useful for nonlinear switching functionalities,
and is distinctly different from the quantum limit in which two
particles already delocalize under weak interactions.

\begin{figure}
\includegraphics[width=8cm]{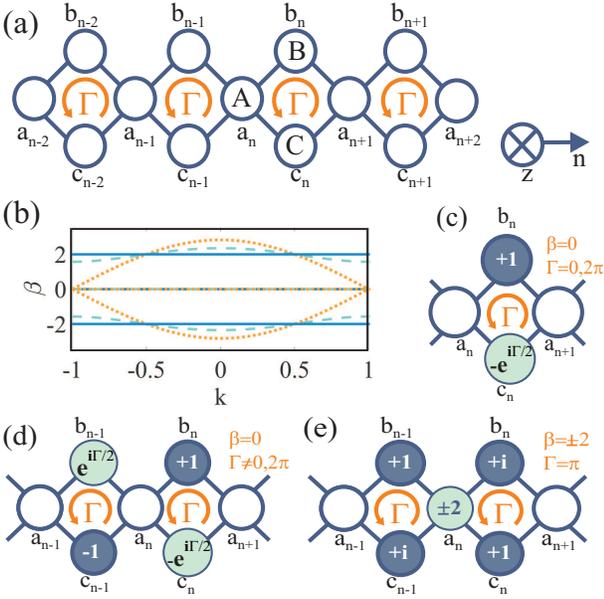}
\caption{(a) Schematic representation of diamond lattice
(characterized by bipartite symmetry) made of coupled optical
waveguides. A synthetic magnetic flux $\Gamma$ is applied in each
plaquette. (b) Band structure for: $\Gamma=0, 2\pi$ (dotted orange
line), $\Gamma=3\pi/4$ (dashed turquoise line) and $\Gamma=\pi$
(blue solid lines). (c-e) Geometry-induced $\beta_{FB} = 0$
flatband modes for different flux strengths (c,d) and additional
flatband modes ($\beta_{FB}=\pm 2$) originating due to the AB
caging (e).} \label{fig1}
\end{figure}

{\it Model.} We consider light propagation in the quasi-1D diamond chain
lattice with synthetically introduced
magnetic flux. The diamond lattice has bipartite
symmetry~\cite{bipartite} with three sites per unit cell: $A, B$
and $C$ as shown in Fig. \ref{fig1}(a). The $A$ sites are fourfold
connected with the nearest neighbors, forming bottlenecks, while
$B$ and $C$ sites make twofold connections with surrounding sites.
Evolution of the optical field $\psi_n = (a_n,b_n,c_n)$ in the presence of on-site nonlinearity is governed by
the discrete nonlinear Schr\"{o}dinger equation,
\begin{eqnarray}
 i \partial_z a_n &=& b_n e^{-i \Gamma/2}+b_{n-1}+c_n+c_{n-1} e^{-i \Gamma/2}
  -g |a_n|^2 a_n, \nonumber  \\
 i \partial_z b_n &=& a_n e^{i \Gamma/2}+a_{n+1} - g |b_n|^2 b_n, \nonumber  \\
 i \partial_z c_n &=& a_n+a_{n+1} e^{i \Gamma/2}-g |c_n|^2 c_n.
 \label{model}
 \end{eqnarray}
Here $z$ is the propagation distance, $g$ is the nonlinearity strength, $\Gamma$ is the flux~\cite{vidal1,Longhi}, $n$ is the unit cell index, and we have normalized the coupling to unity without loss of generality. Experimentally, the flux $\Gamma$ can be
realized either through sinusoidal modulation of refractive index
of waveguides along $z$ direction or by implementation of an
auxiliary waveguide with carefully chosen refractive index in
between two sites~\cite{Experiment,expzamajt,Longhi}. Total beam
power $P = \sum_n (|a_n|^2+|b_n|^2+|c_n|^2)$ and the Hamiltonian are
conserved
quantities. 

In the linear limit ($g=0$), the Bloch wave eigenmodes are
$\{a_n,b_n,c_n\}=(A,B,C)\exp(-i \beta z+i k n)$, where $\beta$ and
$k$ denote the propagation constant and Bloch wave number
respectively. After its substitution into Eq.~\eqref{model} we
obtain the following dispersion relations:
\begin{eqnarray}
\beta_{FB}=0 \nonumber, \quad
\beta_{\pm}= \pm 2 \kappa
\sqrt{1+\cos(\Gamma/2)\cos(k-\Gamma/2)}. \label{DR}
\end{eqnarray}
The $\beta_{FB}=0$ band is completely flat, regardless of the flux $\Gamma$, and is a consequence of the diamond network's topology (plaquettes coupled via bottleneck sites). The other bands are in general $k$ dependent, and only become flat in the AB cage limit $\Gamma = \pi$.

When $\Gamma = 0$ or $2\pi$ the flatband touches two surrounding
dispersive bands at the Brillouin zone edge, as shown in
Fig.~\ref{fig1}(b). Fig.~\ref{fig1}(c) illustrates the fundamental
CLS, which occupies two sites and is localized to a singe unit
cell. In general, when $\Gamma\neq 0, 2\pi$ the $\beta_{FB}=0$
flatband is separated from the others by a gap, and the
fundamental CLS occupying four sites is not orthogonal with
neighboring CLSs, see Fig.~\ref{fig1}(d)~\cite{ibrahim}.
Meanwhile, in the AB cage limit the additional flatbands at $\beta
= \pm 2\kappa$ host five-site CLS that also excite one of the
bottleneck sites [Fig.~\ref{fig1}(e)].

{\it Linear stability of nonlinear CLS.} We start by analyzing how
nonlinearity can lead to instability of the CLS, and potentially
induce transport. Our focus is on the $\beta_{FB}=0$ CLS, which
exists and can be continued as a nonlinear CLS regardless of the
synthetic magnetic flux strength. The general behaviour of
perturbed CLSs can be related to the eigenvalue (EV) spectra which
are obtained via the linear stability analysis (LSA) by applying a
small perturbation $p_n$ to the CLS profile $\psi_n$ and
linearizing the equations of motion
Eq.~\eqref{model}~\cite{Carlo,Belicev}. The eigenvalues $\lambda$
of the linearized equations of motion characterize the initial
stage of instability development of CLS.

\begin{figure}
\includegraphics[width=8cm]{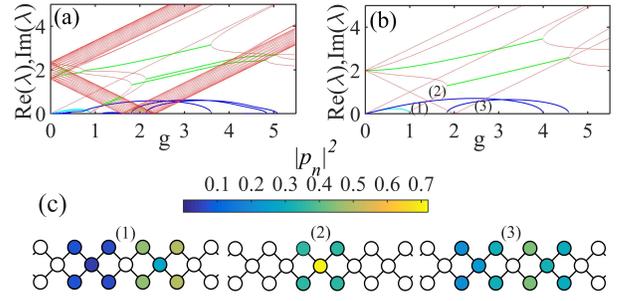}\\
\caption{Linear instability spectrum of nonlinear CLS. (a,b)
Positive frequency part of full linear perturbation eigenvalue
(EV) spectrum for (a) $\Gamma = 3\pi/4$ and (b) $\Gamma = \pi$.
Pure real EVs (cyan) and nonzero real parts of complex EVs (blue)
mark regions of CLS linear instability. Pure imaginary EVs and
imaginary parts of complex EVs are shown in orange and green,
respectively. (c) Characteristic eigenmode profiles of perturbed
CLS corresponding to three EV branches (1), (2) and (3) from (b).
Size of the system is $N=21$.} \label{fig2}
\end{figure}

When $\Gamma\neq 0, 2\pi$, nonlinearity can induce coupling between
neighboring non-orthogonal CLSs. This can be identified in LSA
spectrum via pure real EVs (exponential instabilities) in Fig.
\ref{fig2}. The perturbation eigenmodes are compact and have vanishing tails in the AB cage limit ($\Gamma = \pi$), and are localized with exponential tails for other values of $\Gamma$. This source of instability only induces significant coupling between the very first
neighboring CLSs originating from the same submanifold (same flatband).
It cannot induce longer range spreading.

Typical spectrum of perturbed CLSs emerging from single gapped flatband
located at $\beta_{FB}=0$ is depicted in Fig.~\ref{fig2}(a) for the case $\Gamma=3\pi/4$. Even for small nonlinearity strengths $g$,
the pure real EV branch is accompanied by quartets of complex EVs
with nonzero real parts, indicating oscillatory
instabilities arising due to coupling between the dispersive bands and the CLS.
Since the dispersive states are delocalized,
this mixing can induce spreading of energy through the entire
lattice~\cite{Carlo}. Similar scenario is obtained for $\Gamma=0, 2\pi$.

On the other hand, for $\Gamma=\pi$ the pure real EV branches
simultaneously occur with the complex EVs quartets coming from the
flatbands characterized with $\beta_{FB}=\pm 2$ [Fig.~\ref{fig2}(b)],
whose eigenmodes are also compact and localized. The corresponding
instability eigenmodes share this compact localization, but occupy more sites than the nonlinear CLS, as shown in Fig.~\ref{fig2}(c). Now, the complex EV
branches can be associated with oscillatory instabilities
developed due to the mutual interactions among CLS components
arising from different flatbands. Such oscillations involving multiple
CLS are also observable in the linear case (see single ``B'' site
excitation in Ref.~\cite{expzamajt}) and are specific to AB cages.

\begin{figure}
\includegraphics[width=8cm]{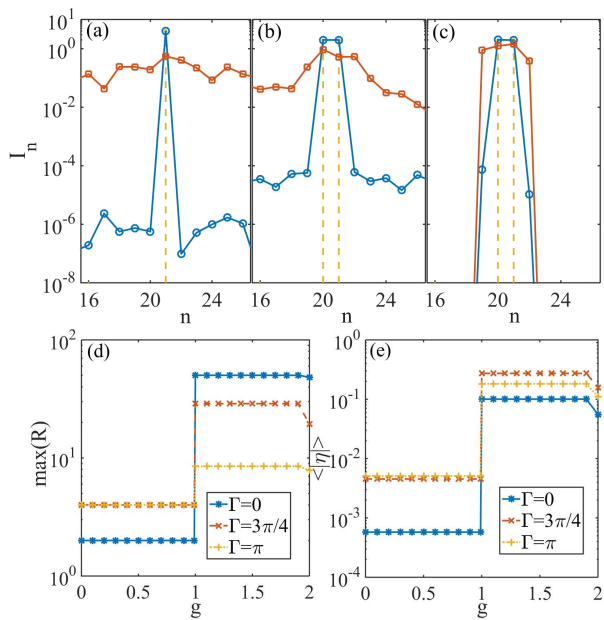}\\
\caption{Nonlinear delocalization and symmetry-breaking of the
CLS. (a-c) Intensity profiles $I_n=|a_n|^2+|b_n|^2+|c_n|^2$ at
$z=10\pi$ for: $\Gamma=0$ (a), $\Gamma=3\pi/4$ (b) and
$\Gamma=\pi$ (c). Cases when $g=0.5$ and $g=1$ are depicted with
blue circle and red square symbol lines, respectively. Vertical
dashed yellow lines mark the cell position of the input CLS. (d)
Spreading of the CLS measured via the maximal value of the
participation ratio $R$ within the propagation length $z=10\pi$.
(e) $z$-average of the normalized leg imbalance $\eta$.}
\label{fig3}
\end{figure}

{\it Propagation dynamics and symmetry breaking.} Above LSA
describes the initial dynamics of a perturbed CLS, assuming all
other modes remain weakly excited. This assumption is typically
satisfied for resonant interactions between nonlinear localized
modes and continua of low amplitude dispersive waves, because the
latter propagate away from the localized mode, stop interacting
with it, and thereby preserve their low amplitude. This argument
fails for AB cages because all low amplitude modes are strictly
localized; if linear instabilities exist, the unstable mode
amplitudes will grow exponentially until nonlinear corrections
become important. What then happens? This question cannot be
resolved by the LSA and must be tackled using numerical
simulations of the propagation dynamics.

Our central result, based on direct simulations of Eq.~\eqref{model} taking CLS with random weak (5\%) perturbations as the initial condition, is that the critical value $g=1$
represents a bifurcation point, beyond which nonlinear symmetry breaking of CLS
occurs. This particular value of nonlinear parameter $g$ separates
a weak instability regime from a strong instability
regime, regardless of the value of the synthetic magnetic flux $\Gamma$, illustrated by the examples in Fig.~\ref{fig3}(a,b,c). For even larger values of $g$ (in the regions where LSA indicates stability), we observe a second transition to conventional self-trapping behaviour for all values of $\Gamma$.

To characterize the transition at $g=1$, we compute the normalized
leg imbalance $\eta = \sum_n (|b_n|^2 - |c_n|^2)/P$ and the
participation ratio $R = P^2/\sum_n (|a_n|^4 + |b_n|^4 +
|c_n|^4)$. The former vanishes for all single band excitations;
nonzero values indicate significant nonlinear interband coupling,
enabling intensity-dependent phase shifts that break the AB cage
by spoiling the destructive interference at the bottleneck sites.
The participation ratio measures the number of sites occupied by
the field, quantifying the wavepacket spreading.

\begin{figure}
\includegraphics[width=8cm]{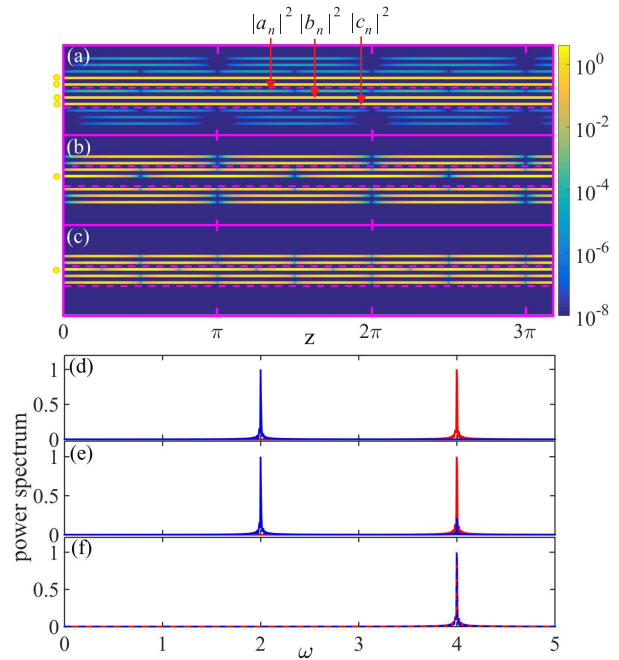}\\
\caption{Periodic dynamics of $\Gamma = \pi$ compact breathing
modes in the weak instability regime $g=0.5$. (a-c) Evolution of
compact breathing modes when initial excitation condition is: (a)
CLS from $\beta_{FB} = 0$ submanifold, (b) single B site and (c)
single A site. Dashed purple lines mark edges of central unit
cell, while yellow circles on the left denote excited sites at $z
= 0$. All plots contain same number of cells. (d-f) Power spectra
of the breathers in (a-c) obtained from Fourier transform of the
site intensities $|a_n(z)|^2$ (red) and $|b_n(a)|^2$ (blue) in the
central unit cell.} \label{fig4}
\end{figure}

As long as $g < 1$, the initially perturbed CLS evolves into a
periodic breather, symmetric with respect to the A site and
keeping almost of all of the total power within the
initially-excited nonlinear mode. Moreover, the dynamics are
confined to the symmetric subspace, i.e. $|b_n| \approx |c_n|$ for
all $z$. Figs. 3(d,e) show sudden increases of $R$ and $\eta$
occur at the bifurcation point $g = 1$. In this strong instability
regime, the breather loses stability and spreading becomes
significant (red symbol lines in Figs.~\ref{fig3}(a-c)). Although
not affecting the precise bifurcation point, AB caging has an
impact on $R$, which is smaller compared to the non-AB cage
networks with $\Gamma \neq \pi$ due to the absence of delocalized
linear modes.

This transition is not specific to the nonlinear CLS. We observe
numerically similar nonlinear transitions for single site
excitations in the AB cage limit $\Gamma = \pi$.
Fig.~\ref{fig4}(a-c) shows representative examples of the weak
instability regime for the CLS as well as B and A single site
excitations, revealing formation of periodic breathers. The
corresponding power spectra in Fig.~\ref{fig4}(d-f) have sharp
peaks corresponding the energy difference between the flatbands.
In particular, the single A site excitation corresponds to a
superposition of eigenmodes of the $\beta_{\pm} = \pm 2\kappa$
bands generating the breather oscillation frequency $\omega = 4$
(Fig. \ref{fig4}(f)); there is no coupling into the $\beta_{FB} =
0$ flatband and the symmetry between the two legs $\eta \approx 0$
is preserved.

\begin{figure}
\includegraphics[width=8cm]{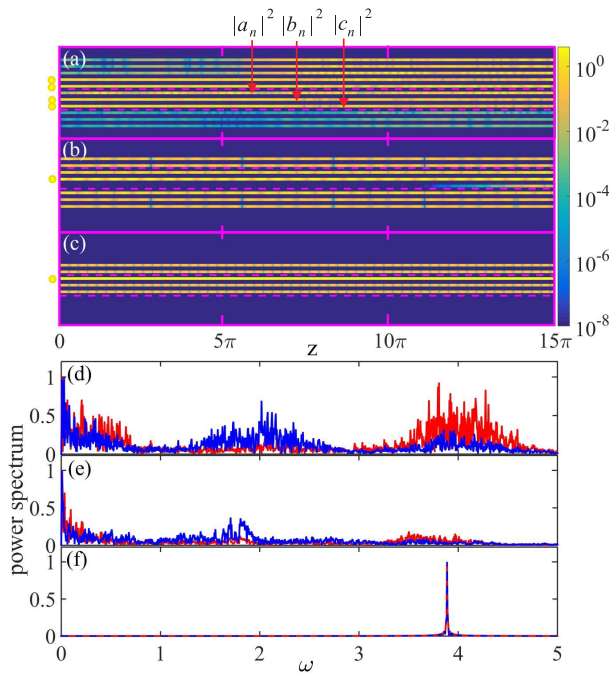}\\
\caption{Dynamics in the strong instability regime $g = 1.9$ for
the three different excitation conditions of Fig.~\ref{fig4}.
(a-c) $z$ evolution exhibiting emergence of symmetry-broken
profiles for the CLS and B excitations (a,b), while the A
excitation remains symmetric. (d-f) Corresponding power spectra of
the intensities in the central unit cell.} \label{fig5}
\end{figure}

In the strong instability regime, a significant fraction of the power is coupled into other modes and the dynamics become aperiodic for the CLS and B site excitations, as shown in Fig.~\ref{fig5}(a,b,d,e). On the other hand, for $1 < g < 2$ the A site excitation forms a different kind of five-site breather, with intensity oscillation frequency $\omega \approx 3.89$ independent of $g$. For $g\ge 2$ the system enters the self-trapping regime, with energy becoming pinned at the initially-excited site. Therefore, an excitation of bottleneck sites remains symmetric regardless of the nonlinearity strength due to the onset of self-trapping, whereas excitations of the legs can undergo the nonlinear symmetry breaking.

{\it Conclusion.} The diamond network with properly tuned
synthetic magnetic flux supports light localization analogous to
AB caging in electronic systems. In the linear limit, destructive
interference at bottleneck sites completely suppresses the
spreading of localized excitations, leading to the existence of
compact localized eigenstates. We showed that such eigenstates can
persist as stable periodic breathers for sufficiently weak
nonlinearities, before undergoing a sudden symmetry-breaking
transition at a critical nonlinearity strength. This symmetry
breaking enables strong spreading of energy to neighboring cells,
breaking the AB caging. We confirmed these statements using linear
stability analysis and direct beam propagation simulations,
observing that the symmetry-breaking transition is not sensitive
to the precise strength of the effective magnetic flux. This
transition occurring at finite nonlinearity strength is notably
distinct from previous studies focusing on the quantum limit,
where all two-particle eigenstates become delocalised for
arbitrarily weak interaction strengths.

Nonlinear breathers and symmetry breaking in AB cages are readily
observable in femtosecond laser-written arrays similar to those in
Refs.~\cite{expzamajt,Experiment}. Previous experiments reported
an effective nonlinear coefficient of $\gamma = 1.7$
cm$^{-1}$MW$^{-1}$ with probe beam powers up to $P = 4$
MW~\cite{heinrich}. Meanwhile, the AB cage experiment of
Ref.~\cite{expzamajt} reported an effective coupling strength of
$\kappa = 0.85$ cm$^{-1}$ with a propagation length $L = 10$ cm,
corresponding to dimensionless propagation length $\kappa L = 8.5$
with a normalized nonlinear coefficients $g = P \gamma / \kappa$
up to 8, sufficient to observe the three nonlinear regimes of weak
instability, strong instability, and self-trapping. As the
underlying mechanisms are general, relying only on the interplay
between magnetic flux and nonlinear phase shifts at bottlenecks in
the AB cage network, we anticipate it can be generalized to
two-dimensional AB cage structures such as the dice
lattice~\cite{Vidal,moller}, and other nonlinear platforms such as
exciton-polariton condensates~\cite{polariton_FB}.

\begin{acknowledgments}  We acknowledge support from the
Ministry of Education, Science and Technological Development of
the Republic of Serbia (Project No. III 45010) and the Institute for Basic Science in Korea (IBS-R024-Y1).
\end{acknowledgments}

\end{document}